\documentclass[12pt]{article}
\usepackage{graphicx}
\usepackage[hang,small,bf]{caption}
\let\Large=\large
\newcommand{\be}[3]{\begin{equation}  \label{#1#2#3}}    

\newcommand{\ee}{ \end{equation}}
\newcommand{\ba}{\begin{array}}
\newcommand{\ea}{\end{array}}

\renewcommand{\arraystretch}{1.8}

\begin{document}

\begin{flushright}
CAMS-98/6\\
HUB-EP-98/69 \\
UPR-820-T\\
hep-th/9810227  
\end{flushright}
\vspace{1cm}
\begin{center}
\baselineskip=16pt

\centerline{\Large \bf Non-Extreme Black Holes of  Five Dimensional}

\bigskip

\centerline{\Large \bf  N=2 AdS Supergravity}

\vspace{2truecm}
\centerline{\bf 
K. Behrndt$^a$\footnote{behrndt@physik.hu-berlin.de}\ , \ 
M. Cveti{\v c} $^{b}$\footnote{cvetic@cvetic.hep.upenn.edu}
 and  \ W. A. Sabra$^{c}$\footnote{ws00@aub.edu.lb}}
\vspace{.5truecm}
{\em
\centerline{$^a$Humboldt-Univ., Institut f\"ur Physik, 
Invalidenstra\ss e 110, 10115 Berlin, Germany}
\centerline{$^b$ Dept.\ of Physics and Astron., Univ.\  of
Pennsylvania, Philadelphia, PA 19104, U.S.A.}
\centerline{$^c$Center for Advanced Mathematical Sciences, 
American University of
 Beirut, Lebanon.}}
\end{center}

\vskip 1 cm
\begin{abstract}
We derive and analyse the full set of equations of motion for
non-extreme static black holes (including examples with the spatial
curvatures $k=-1$ and $k=0$) in D=5 $N$=2 gauged supergravity by
employing the techniques of ``very special geometry''. These solutions
turn out to differ from those in the ungauged supergravity only in the
non-extremality function, which has an additional term (proportional
to the gauge coupling $g$), responsible for the appearance of naked
singularities in the BPS-saturated limit.  We derive an explicit
solution for the $STU$ model of gauged supergravity which is 
incidentally also a solution of D=5 $N$=4 and $N$=8 gauged supergravity.
This solution is specified by three charges, the asymptotic negative
cosmological constant (minimum of the potential) and a non-extremality
parameter. While its BPS-saturated limit has a naked singularity, we
find a lower bound on the non-extremality parameter (or
equivalently on the ADM mass) for which the non-extreme solutions are
regular. When this bound is saturated the extreme
(non-supersymmetric) solution has zero Hawking temperature and finite entropy.
Analogous
qualitative features  are expected to emerge for black hole solutions
in $D=4$ gauged supergravity as well.
\end{abstract}
\bigskip


\newpage


\section{Introduction}


Recently, there has been renewed interest in gauged supergravity
theories in various dimensions. It is motivated by the fact that the
ground state of these theories is anti-deSitter (AdS) space-time and
thus they may have implications for the recently proposed AdS/CFT
correspondence~\cite{160,170,180}, which implies an equivalence of
Type IIB string theory (or M-theory) on anti-deSitter (AdS) space-time
and the conformal field theory (CFT) on the boundary of this
space.

Specifically, Type IIB string theory on $AdS_5\times S_5$ is conjectured
\cite{160} to be dual to D=4 $N=4$ superconformal Yang-Mills
theory in the infinite t'Hooft coupling limit $g_{YM}^2 N \rightarrow
\infty$.
It is of special interest to address cases with less
than 32 conserved supercharges and thus lower or no supersymmetry, in
order to shed light on the nature of the correspondence
there. Supergravity vacua with less supersymmetry may have an
interpretation on the CFT side as an expansion of the theory around
non-zero vacuum expectation value of certain operators. (Solutions
with no supersymmetry could also be viewed as excitations above the
ground state of the theory.)

One set of non-trivial gravitational backgrounds which preserve only
part of the symmetry are BPS-saturated solutions, e.g., BPS-saturated
black holes.  Unfortunately, D=5 static BPS-saturated black
holes~\cite{130} of gauged supergravity have naked singularities and
thus, their singular geometry indicates ill defined properties of the
theory at small distances on the  gravity side~\footnote{BPS-saturated
topological black holes in gauged supergravity, also with naked singularities, 
 were obtained in \cite{135,136}.}.


The purpose of this paper is to explore non-extreme static black hole
solutions of D=5 gauged supergravity.  One of the motivation of this
study is to shed light on the geometry of these solutions, in
particular their singularity structure.  In particular we would like
to explore the range of ADM mass parameters for which the horizons are
present and in turn determine their thermodynamic features. These
features could potentially provide an insight into dynamics
of Yang-Mills theories with broken supersymmetry.

Since the standard D=5 black holes have a spherical $S_3$-symmetry,
these black holes may act as a gravitational background for the dual
Yang-Mills theory with the global geometry of $R \times S_3$.  (It may
be important to replace the $S_3$ by
a more general three-dimensional space with constant curvature $k$,
i.e. along with the ordinary static black holes with $k=1$, static
solutions with $k=-1$ and $k=0$ may also be of interest.)
Interestingly, now the (charged) solutions have an extreme limit (with a
vanishing Hawking temperature) which does {\em not} coincide with the
BPS-saturated limit. So, they may serve as a non-supersymmetric
gravity background at zero temperature.  This situation is similar to
the four-dimensional case, where charged black holes of gauged
supergravity also allow for a zero-temperature non-supersymmetric
limit \cite{220}.

Within this more general setting we address such static black holes,
with $k=\pm 1,\ 0$.  After briefly reviewing D=5 $N=2$ gauged
supergravity theory in Section 2 we derive the equations of motion for
the specific field Ans\"atze in Section 3.  (Note that a subclass of
solutions of $N=2$ supergravity are actually also solutions of
supergravity theories with more, i.e. $N=4$ or $N=8$ supersymmetries.)
In Section 4 we write an explicit solution for the case of a special
prepotential, which is a gauged version of the three charge solution
of ordinary $N=4,8$ supergravity ~\cite{190}, \cite{260}.  (For
equal charges a gauged solution has been discussed in \cite{230}.) 
 For $k=1$ solutions we specifically identify the lower bound on the
non-extremality parameter (or equivalently the ADM mass) which ensures
that these solutions have regular horizons and further discuss their
thermodynamic features.  


\section{D=5  $N$=2  gauged supergravity}


In the context of M-theory, the theory of five-dimensional $N=2$
supergravity coupled to abelian vector supermultiplets arise by
compactifying eleven-dimensional supergravity, the low-energy theory
of M-theory, on a Calabi-Yau three-folds \cite{99,100}. The massless
spectrum of the theory contains $(h_{(1,1)}-1)$ vector multiplets with
real scalar components, and thus $h_{(1,1)}$ vector bosons (the
additional vector boson is the graviphoton). The theory also contains
$h_{(2,1)}+1$ hypermultiplets, where $h_{(1,1)}$ and $h_{(2,1)},$ are
the Calabi-Yau Hodge numbers.  The anti-de Sitter supergravity can be
obtained by gauging the $U(1)$ subgroup of the $SU(2)$ automorphism
group of the $N=2$ supersymmetry algebra, which breaks $SU(2)$ down to
the $U(1)$ group. The gauging is achieved by introducing a linear
combination of the abelian vector fields already present in the
ungauged theory, i.e. $A_\mu=V_IA_\mu^I$, with a coupling constant
$g$. The coupling of the fermi-fields to the $U(1)$ vector field
breaks supersymmetry, and in order to preserve $N=2$ supersymmetry,
one has to introduce gauge-invariant $g$-dependent terms.  In a
bosonic background, these additional terms give a scalar potential
\cite{110}.

The bosonic part of the effective gauged supersymmetric $N=2$
Lagrangian which describes the coupling of vector multiplets to
supergravity is given by
\begin{eqnarray}
e^{-1} \mathcal{L} &=& {\frac{1}{2}} R + g^2V - 
{\frac{1}{4}} G_{IJ} F_{\mu\nu}
{}^I F^{\mu\nu J}-{\frac{1}{2}} {\cal G}_{ij} \partial_{\mu} 
\phi^i \partial^\mu
\phi^j +{\frac{e^{-1}}{48}} \epsilon^{\mu\nu\rho\sigma\lambda} C_{IJK}
F_{\mu\nu}^IF_{\rho\sigma}^JA_\lambda^k \ ,   \nonumber \\
\label{action}
\end{eqnarray}
with the space-time indices $(\mu,\nu)=0,1,\cdots, 4$ have $(-1, 1,
\cdots , 1)$ signature, $R$ is the scalar curvature, $F_{\mu\nu}^I$
are the Abelian field-strength tensors and $e=\sqrt{-g}$ is the
determinant of the F\"unfbein $e_m^{\ a}$, $V$ is the potential given
by
\be020
V(X)= V_I V_J\Big( 6 X^I X^J - {9 \over 2} 
{\cal G}^{ij} \partial_iX^I \partial_j X^J \Big)
\ee
where $X^I$ represent the real scalar fields which have to satisfy the
constraint 
\be030
{\cal V} = {1 \over 6} C_{IJK} X^I X^J X^K =1 \ .
\ee
Also:
\be040
G_{IJ} = -{1 \over 2} \partial_I \partial_J \log {\cal V} \Big|_{{\cal V} = 1}
\quad , \quad
{\cal G}_{ij} = \partial_i X^I \partial_j X^J G_{IJ} \Big|_{{\cal V} = 1} \ ,
\ee
where $\partial_i$ refers to a partial derivative with respect to the
scalar field $\phi^i$.  The physical quantities in (\ref{action}) can
all be expressed in terms of the homogeneous cubic polynomial ${\cal
V}$ which defines a \lq\lq very special geometry'' \cite{120}.

Further useful relations are
\be050
\partial_iX_I = - {2 \over 3} G_{IJ} \partial_i X^J
\quad , \quad
X_I = {2 \over 3} G_{IJ} X^J \ .
\ee
It is worth pointing out that for Calabi-Yau compactification, ${\cal
V}$ is the intersection form, $X^I$ and $X_I={1\over6}C_{IJK}X^JX^K$
correspond to the size of the two- and four-cycles and $C_{IJK}$ are
the intersection numbers of the Calabi-Yau threefold.

Using the relationship (which can be proven within techniques of very
special geometry)
\be060
{\cal G}^{ij} \partial_j X^I \partial_j X^J = G^{IJ} - 
{2 \over 3} X^I X^J \ , 
\ee
the potential can also be written as
\be070
V(X)= 9\,V_I V_J\Big( X^I X^J - {1 \over 2} G^{IJ} \Big)\ .
\ee


\section{The Equations Of Motion}


Before turning to the equations of motion we discuss the Ansatz for
the non-extreme solution. For the standard static black holes the
geometry at a given radius is $R \times S_3$ where the three-sphere
becomes the horizon at $r=r_H$. However in the context of CFT/AdS
correspondence it may be useful to discuss a more general class of
solutions, where the $S_3$ is replaced by a three-dimensional space
with constant curvature $k$. (In D=4 this replacement has been
discussed for topological black holes \cite{200}, where the horizon is a
genus $g$ Riemann surface.) To be specific we consider a hypersurface
given by the equation
\be080
X_1^2 + X_2^2 +X_3^3 + {k \over {|k|}} X_4^2 = {1 \over k}\ , 
\ee
i.e.\ for $k=1$ it is a $S_3$, for $k=-1$ it is a pseudo-sphere and
for $k=0$ it is a flat space. Introducing angular coordinates, the
metric becomes
\be090
d\Omega_{3,k} = d\chi^2 + \Big({\sin \sqrt{k} \chi \over \sqrt{k}}\Big)^2
\Big(d\theta^2 + \sin^2\theta d\phi^2\Big) \ .
\ee
Taking this spherical part, our Ansatz for the metric and the scalars
reads
\be100
ds^2 = - e^{-4U} f dt^2 + e^{2U}\left( {dr^2 \over f} + 
	r^2 d\Omega_{3,k} \right) \qquad , \qquad
X_I = {1 \over 3} e^{-2U}  H_I \ .
\ee
Motivated by the form of $f$ in the case of the BPS-saturated solution
\cite{130}, i.e. $f_{BPS}=1+g^2r^2e^{6U}$, and the form for the
non-extreme solutions in the ungauged case, i.e. $f=k-\mu/r^2$, we
take the following Ansatz for the function $f$:
\be170
f = k - {\mu \over r^2} + g^2 r^2 e^{6U}\ , 
\ee
where $\mu$ is the non-extremality parameter.  On the other hand the
Ansatz for the $U$-function and the scalars $X^I$ remains the same as
for the extreme case and it is given by a set of harmonic
functions:
\be110
H_I = h_I + {q_I \over r^2}\ .
\ee
Note that the choice to express the above Ans\"atze (\ref{170}) for
$f$ and (\ref{100}) for $X_I$ in terms of {\it harmonic functions}
$H_I$ (\ref{110}) is special and corresponds to solutions with a
special form of the prepotential ${\cal V}$ (\ref{030}),
i.e. ``toroidal''-type compactifications. (Note that the discussion of
the Einstein equations in Section 3.2 relies heavily on this form of
the Ans\"atze.) In general $H_I$ need not be harmonic and thus the
derived equations of motion for the harmonic form of $H_I$ need not
have a consistent solution (See a discussion at the end of Section 3).


\subsection{Gauge field equations}


In solving the equations of motion we start with the gauge field
equation, which reads
\be120
{1 \over \sqrt{g}} \, \partial_{\mu} 
\left( \sqrt{g} \, G_{IJ} F^{J\, \mu\nu} \right)
\sim \epsilon^{\nu \alpha \beta \gamma \lambda} F_{\alpha \beta}^J
F_{\gamma \lambda}^K C_{IJK} \ .
\ee
Since we are considering only electrically charged and non-rotating
solutions ($F_{mn} =0$, $(m,n) =(1\cdots4)$) the right-hand-side (rhs)
vanishes. Thus,
\be130
{1 \over \sqrt{g}} \, \partial_{r} \left( \sqrt{g} \, G_{IJ} \, g^{00} \,g^{rr}
	F^{J}_{r0} \right) = {1\over r^3} \, 
        \partial_{r} \left( r^3 \, e^{4U} \, G_{IJ} 
	F^{J}_{r0} \right) = 0\ , 
\ee
which is solved by
\be140
F_{r0}^J = -{\sqrt{k} \over 2} e^{-4U} G^{JI} \partial_r \tilde H_I \ ,
\ee
where 
\be145
\tilde H_I=1 +{{\tilde q}_I\over r^2}\ ,
\ee 
is a new set of harmonic functions with parameters ${\tilde q}_I$
corresponding to the physical electric charges.  Note that in the
extreme limit the ${\tilde H}_I$'s turn out to coincide with the
$H_I$'s, introduced in (\ref{110}).  We have chosen the coefficient in
front of the rhs of ({\ref{140}) in order to get the known extreme
solution with $k=1$. The appearance of $\sqrt{k}$ will be motivated
below, namely, the Einstein equations can be solved in the
extreme-case and for $k=0, \ -1$ only when the coefficient on the rhs
of (\ref{140}) is chosen to be proportional to
$\sqrt{k}$. (c.f. eq. (\ref{260})).

The appearance of the harmonic function in (\ref{140}) seems to
indicate that the generalization to a multi-center solution is
straightforward.  This, however, is not the case.  The Bianchi
identity is solved only, if the solution depends on $r$ only, or
otherwise for $\tilde H_I = H_I$, which corresponds the extreme case.


\subsection{Einstein equations}


When expressed only in terms of the Ricci tensor the Einstein equation
becomes
\be150
R_{\mu\nu} =  \left(F^2_{\mu\nu} - {1 \over 6} g_{\mu\nu} F^2 \right)
   + \partial_{\mu} X^I \partial_{\nu} X^J \, G_{IJ} 
   - {2 \over 3} \, g^2 \, V(X) g_{\mu\nu} \ ,
\ee
with $F^2_{\mu\nu} \equiv G_{IJ} F_{\mu\rho}^{I} F^{J}_{\nu\lambda} \,
g^{\rho\lambda}$ and we have used $\partial_{\mu} \phi^i
\partial_{\nu} \phi^j {\cal G}_{ij} = \partial_{\mu} X^I
\partial_{\nu} X^J G_{IJ}$.  First we consider the combination that
determines the $f$-function
\be160
R_0^{\ 0} + 2 R_{\theta}^{\ \theta} = - 2\, g^2 \, V(X) \ .
\ee
Calculating the Ricci tensor for our metric Ansatz, one finds 
\be180
g^{-2} (R_0^{\ 0} + 2 R_{\theta}^{\ \theta}) \equiv 
-e^{-2U} \left[ 12 \, e^{6U} + {11 \over 2} r \, (e^{6U})' + {r^2 \over 2} 
\, (e^{6U})'' \right] 
\ .
\ee
(primes refer to derivatives with respect to $r$). Therefore,
the dependence on $\mu$ and the spatial curvature $k$ drop out, thus
confirming the correct dependence of $f$ (\ref{170}) on $\mu$.  Note
that in the special case of $U=0$ the scalars are constant and the
potential becomes constant $V=6$, corresponding to the $AdS$ vacuum.

{From} (\ref{100}) one obtains that $e^{2U} = {1 \over 3} X^I H_I$
(recall $X^I X_I = 1$) and using the relation $H_I \partial_i X^I \sim
X_I \partial_i X^I = 0$ we find
\be190
(e^{2U})' = {1 \over r} \Big[ {2 \over 3 } (Xh) - 2 e^{2U} \Big]\ ,
\ee
with $(Xh) \equiv X^I h_I$. Similarly,
\be200
(X^I)' = e^{-2U} {1 \over r} \Big[ {2 \over 3} X^I (Xh) - G^{IJ} h_J \Big] \ .
\ee
Using these relations we get
\be210
g^{-2} (R_0^{\ 0} + 2 R_{\theta}^{\ \theta}) = 
- e^{-2U} \left[ 2 (Xh)^2 - |h|^2 \right] = - 2 h_I h_J ( X^I X^J - 
{1\over 2} G^{IJ})\ ,
\ee
where $|h|^2 \equiv G^{IJ} h_I h_J$.  Comparing the form of the
potential $V$ given in (\ref{070}) with the rhs of (\ref{210}), we
precisely reproduce the equation (\ref{160}), providing the
following relationship between the constant parts $h_I$ of the
harmonic functions $H_I$ (\ref{110}) and the expansion coefficients
for the gauge field $A_{\mu} = V_I A_{\mu}^I$ is satisfied:
\be220
h_I = 3 V_I \ .
\ee
(The choice $V_I={1\over 3}$ ensures the canonical normalization for
the harmonic functions $H_I$ (\ref{110}) with $h_I=1$.)  Thus, we have
verified the equation (\ref{160}) is satisfied with the Ans\"atze
(\ref{170}) for $f$ and (\ref{220}) for the constant part of the
harmonic functions (\ref{110}).  We can thus use (\ref{180}) to bring
the components of the Ricci tensor in the following simpler form:
\be230
\ba{rcl}
R_0^{\ 0} &=& 2 e^{-2U} \Big(k - {\mu \over r^2}\Big) \Big( U'' + {3 \over r} 
  U' \Big) + 2 e^{-2U}  \Big(k - {\mu \over r^2}\Big)' U' - 
  {2\over 3} g^2 V \ ,
	 \\
R_i^{\ i} &=& - e^{-2U} \Big(k - {\mu \over r^2}\Big) 
	\Big( U'' + {3 \over r} U' \Big) 
	- e^{-2U}  \Big(k - {\mu \over r^2}\Big)' U' 
	- {2\over 3} g^2 V \ , \\ 
R_r^{\ r} &=& - e^{-2U} \Big(k - {\mu \over r^2}\Big) 
	\Big( U'' + 6 (U')^2 + {3 \over r} U' \Big) + 
	2 e^{-2U}  \Big(k - {\mu \over r^2}\Big)' U' - {2\over 3} g^2 V  \\
&&	- 3 g^2 r^2 e^{4U} \Big( U'' + 2 (U')^2 + {3 \over r} U'  \Big)\ . 
\ea
\ee
(Note there is no summation over the index $i$, which refers only to
the angular components.)  Interestingly, all the dependence on the
gauge potential ($\propto g^2$) drops out of the Einstein equations
(\ref{150}) and thus these equations are identical to those obtained
for the non-extreme black holes of ungauged $N=2$ supergravity.

As next step we consider the $R_0^{\ 0}$ component of Einstein
equations, which becomes
\be240
R_0^{\ 0} + {2 \over 3} g^2 V = {1 \over 3} F^2  \equiv
	{1 \over 3}\, G_{IJ} F^I_{\mu\nu} F^{J\, \mu\nu} = {k \over 6}
	e^{-6U} G^{IJ} \partial_r \tilde H_I \partial_r \tilde H_J  \ ,
\ee
where we have used $F_{\;0}^{2\ 0} - {1 \over 6} g_{0}^{\ 0} F^2 = {1
\over 3} F^2$ (since we have only electric fields) and we have
inserted the form of the gauge field (\ref{140}). In order to simplify
the analysis we can employ the fact that the form of the $U$ function
is the same as in the extreme case (i.e. when $\tilde H_I = H_I$, $\mu
= 0$), and thus it solves the following equation
\be250
2e^{-2U} \Big( U'' + {3 \over r} U' \Big) =
	{1\over 2} e^{-6U} G^{IJ}\partial_r H_I \partial_r H_J 
	\ .
\ee
Hence, the $R_0^{\ 0}$ equation becomes
\be260
2 e^{-2U}  \Big(k - {\mu \over r^2}\Big)' U' =  
	\textstyle{1\over 6} e^{-6U}G^{IJ} \left[\Big(k - 
{\mu\over {r^2}}\Big) 
	\partial_r H_I \partial_r H_J - k\partial_r \tilde H_I \partial_r 
	\tilde H_J 
\right]\ .
\ee
Introducing $\tilde \mu \equiv {\mu \over k}$ (for $k \ne 0$) in
(\ref{260}) yields:
\be270
2\Big(e^{6U}\Big)'  \; { \tilde \mu \over r^3} =  
	{1 \over 2} e^{2U}G^{IJ} \left[\Big(1 - {\tilde \mu \over r^2}\Big) 
	\partial_r H_I \partial_r H_J - \partial_r \tilde H_I \partial_r 
	\tilde H_J
\right] \ .
\ee
On the other hand for $k=0$ we obtain:
\be275
2\Big(e^{6U}\Big)'  \; {\mu \over r^3} =  
	{1 \over 2} e^{2U}G^{IJ} \left( - {\mu \over r^2}\right) 
	\partial_r H_I \partial_r H_J\ .
\ee
The $R^{\ r}_r$ part of the Einstein equations:
\be280
R_r^{\ r} = {1 \over 3} F^2 + \partial_r X^I \partial^{\, r} X^J G_{IJ} 
	-{2 \over 3} g^2 V \ .
\ee
can be cast in the same form as (\ref{270}). Namely, again using the
relations (\ref{190}) and (\ref{200}) we cast the scalar kinetic part
in the following form:
\be290
\partial_r X^I \partial_r X^J G_{IJ}  g^{rr} = {1 \over r^2} e^{-6U} 
	\Big[-{2 \over 3} (X h)^2 + |h|^2 \Big]
	\Big(k - {\mu \over r^2} + g^2 r^2 e^{6U} \Big)\ , 
\ee
and 
\be300
U'' + 2 (U')^2 + { 3\over r} U' = e^{-4U} {1 \over 3r^2} \Big[ {2\over 3}
	(X h)^2 - |h|^2 \Big] \ .
\ee
Inserting (\ref{290}) and (\ref{300}) in  (\ref{280})  we arrive 
at the same equation as  that for the $R^{\ 0}_0$  
component of the Einstein equations (\ref{270}). 

Due to the symmetry of the $R^{\ i}_i$  (angular) components of the 
Einstein equations,  the rest of $R^{\ i}_i$ components are
redundant,  and thus the   analysis of the  Einstein equations is complete.


\subsection{Scalar equations}


The scalar field equation reads
\be320
{1 \over \sqrt g} \partial_{\mu} \Big( \sqrt g g^{\mu\nu} {\cal G}_{ij}
	\partial_{\nu} \phi^j \Big) = {1 \over 4} \partial_i G_{JK}
	F^{J}_{\mu\nu} F^{K\, \mu\nu} - g^2 \partial_i V(X) \ ,
\ee
where $\partial_i$ refers to a partial derivative with 
respect to the scalar fields $\phi^i$.
The left-hand-side (lhs) of (\ref{320})  becomes
\be330
{e^{-2U} \over r^3 }\partial_r \Big[r^3 (k - {\mu \over r^2} + 
g^2 r^2 e^{6U}) \, {\cal G}_{ij} \partial_r \phi^j \Big] \ .
\ee
We again employ the fact, that the scalar fields  are independent of $k$, $\mu$
and $g$. i.e. they have the same form as in the known extreme case for
$k=1$. Therefore, multiplying the extreme equation (i.e. taking
$\tilde H_I = H_I$ in the field strengths) with $(k-{\mu \over r^2})$
and subtracting it from both sides, only the term $\propto (k-{\mu
\over r})'$ survives and we find
\be340
\ba{rcl}
{e^{-2U}}(k - {\mu \over r^2})' \, {\cal G}_{ij} \partial_r \phi^j  &=& 
{1 \over 4} \partial_i G_{JK} \left[
	F^{J}_{\mu\nu} F^{K\, \mu\nu} - (k - {\mu \over r^2}){1 \over k} 
	F^{J}_{\mu\nu} F^{K\, \mu\nu}\Big|_{\tilde H_I = H_I}\right] \\
&=& \textstyle{1 \over 8} e^{-6U} \partial_i G^{JK}\left[\Big(k - 
      {\mu\over r^2}\Big) 
	\partial_r H_I \partial_r H_J - k\partial_r \tilde H_I \partial_r 
	\tilde H_J
\right] \ .
\ea
\ee
Again, introducing ${\tilde \mu}\equiv{\mu\over k}$ the equation
(\ref{340}) 
takes the
following form:
\be345
2e^{6U}{{\tilde \mu} \over r^3}\,
 {\cal G}_{ij} \partial_r \phi^j  =
\textstyle{1 \over 8} e^{2U} \partial_i G^{JK}\left[\Big(1 - 
{{\tilde \mu}\over r^2}\Big) 
	\partial_r H_I \partial_r H_J - \partial_r \tilde H_I \partial_r 
	\tilde H_J
\right] \ .
\ee
while for $k=0$ eq. (\ref{340}) can be more more conveniently written as:
\be347
2e^{6U}{{\mu} \over r^3} \, {\cal G}_{ij} \partial_r \phi^j  =
\textstyle{1 \over 8} e^{2U} \partial_i G^{JK}\left(- {{\mu}\over r^2}\right) 
	\partial_r H_I \partial_r H_J 
 \ .
\ee

In conclusion, we have analysed all the equations of motion of the
Lagrangian (\ref{action}) with the static Ans\"atze (\ref{100}),
(\ref{110}) for the metric and the scalar fields. The analysis of the
gauge field equations in Subsection 3.1.\ introduced the harmonic
functions $\tilde H_I$ (c.f.  (\ref{140})), the analysis of Einstein
equations in Subsection 3.2 confirmed the Ansatz (\ref{170}) for the
function $f$ in (\ref{100}).  In addition it yielded {\it one}
additional constraint given by equation (\ref{270}).  The study of the
scalar equations in Subsection 3.3 yields one more set of equations
(\ref{345}). Thus, solving these equations will fix the remaining
parameters in the harmonic functions ($H_I, \tilde H_I$).  Note that up
to the replacement of $\mu\to \tilde \mu$ both (\ref{270}) and
(\ref{345}) are the same as in the ungauged case!  


In general, the  equations of motion,
 i.e.\ (\ref{270}) and (\ref{345}) (or
(\ref{275}) and (\ref{347})), cannot be solved in terms of the 
  harmonic function Ans\" atze (\ref{110}), only and 
  one should  instead regard functions $H_I$, which
determine $\phi^i$ and $e^{2U}$, as general functions, not necessarily
harmonic. In this case these equations become coupled second order
differential equations including a dumping term (proportional to the
first derivative) and a potential (coming from the field strengths);
see also \cite{140}.  

On the other hand, for a specific choice of $\cal V$ it is possible to
find an explicit solution in terms of harmonic functions, in
particular for analogs of the ``toroidal''-type compactifications
discussed for the ungauged cases in \cite{150,140}. This is the
three-charge configuration with no self-intersections, i.e.\ only $C_{123}
\neq 0$ in (\ref{030}).  In this case the solution for $U$ and
$\phi^i$ can be expressed in terms of harmonic functions $H_I$, and
the two sets of harmonic functions $H_I$ and $\tilde H_I$ are
proportional to each other by a constant matrix~\cite{140}.  In the
following we will analyse in detail a specific example in this
subclass, the $STU$ model.


\section{Discussion of a special  solution}


As an example we consider the $STU$ model
 which   has only one intersection number
$C_{123}=1$ nonzero.  This model can be embedded into gauged $N=4$ and
$N=8$ supergravity as well.  In the following we shall derive the
explicit solution and its properties. 


\subsection{Solution for the $STU$ model}


This model is given by the prepotential 
\be350
{\cal V} = STU = 1
\ee
Taking $S=X^1$, $T=X^2$ and $U=X^3$ one gets for $e^{6U}$ and the
matrix $G^{IJ}$
\renewcommand{\arraystretch}{1}
\be360
e^{6U} = H_1 H_2 H_3 \qquad , \qquad
G^{IJ} = 2 \left( \ba{ccc} S^2 && \\ & T^2 & \\ && U^2 \ea \right)\ .
\ee
\renewcommand{\arraystretch}{1.8}
Considering $S$ as the dependent field, i.e.\ $S = 1/TU$ we find
\be370
{\cal G}_{ij} = \left( \ba{cc} {1 \over T^2} & {1 \over 2 TU} \\
{1 \over 2 TU} & {1 \over U^2} \ea \right) \quad , \quad
{\cal G}^{ij} = {4 \over 3} \left( \ba{cc} T^2 & - {TU \over 2 } \\
 - {TU \over 2 } & U^2 \ea \right) \ .
\ee
For this case the potential reads (assuming $h_I=(1,1,1)$ and thus
$V_I={1\over 3}$, c.f. (\ref{220})):
\be380
V(T,U) = 2\Big({1 \over U} + {1 \over T} + TU \Big)
\ee
with the minimum $V_{min}(T=U=1) = 6$ which is reached in the
asymptotic vacuum with cosmological constant given by $g^2 V_{min}$.

The  Ans\"atze (\ref{100}), along with  (\ref{140}) and (\ref{170}),
yield the following  explicit form for the fields:
\be390
\ba{l}
ds^2 = - {(H_1 H_2 H_3)^{-2/3}} f dt^2 + (H_1H_2H_3)^{1/3}
 	\bigg(f^{-1} {dr^2} + 
	r^2 d\Omega_{3,k} \bigg) \ ,  \\ 
f = k - {\mu \over r^2} + g^2 r^2 H_1H_2H_3 \quad , \quad
X^I = H_I^{-1}{(H_1 H_2 H_3)^{1/3} } \quad , \quad
F_{r0}^I = -{\sqrt{k} \over 2} (H_I)^{-2 } \partial_r \tilde H_I\ , 
\ea
\ee
where $k$ determines the spatial curvature of $d\Omega_{3,k}$.
Notice, for $k=0$ the gauge fields vanish, but the scalars remain
non-trivial. Finally $(q_I , \tilde q_I)$ are fixed by the equations
(\ref{270}) and (\ref{340}) and one finds
\be392
q_I = \tilde \mu \sinh^2\beta_I \qquad , \qquad 
\tilde q_I = \tilde \mu \sinh \beta_I \cosh\beta_I
\qquad (\tilde \mu\equiv { \mu \over k}) \ .
\ee
which are the same expressions as in the ungauged case (since the
equations are the same).

Note also, that for the extreme case ($\mu=0$) with $k=0$ and equal charges, i.e. 
$\beta_1=\beta_2=\beta_3$ ($H_1$=$H_2$=$H_3$) we
find exactly the $AdS_5$ part of the $D3$-brane!



In the following subsections we turn to the discussion of the global space-time 
structure and  thermodynamics  of these solutions. We will
restrict ourselves to the case of
$k=+1$, only;  the global
structure for $k=-1,\  0$ is very different and will be discussed elsewhere.
\subsection{ADM mass}

In order to determine the ADM mass we will follow a procedure given by
Horowitz/Myers \cite{210} (a generalization of the Nester's
procedure for asymptotically non-flat space-time). First by
defining a new radial coordinate 
\be410
\rho^2\equiv  r^2(H_1H_2H_3)^{1/3}\ , 
\ee
the metric (\ref{390}) can be written as:
\be420
ds^2 = - e^{-2V} dt^2 + e^{2W} d\rho^2 + \rho^2 d\Omega_{3,k}\ , 
\ee
where:
\be425
e^{-2V} = f (H_1H_2H_3)^{-2/3}, \ \  
 e^{2W}=f^{-1}(H_1H_2H_3)^{1/3} \left({{d\, r}\over
{d\, \rho}}\right)^2 \ , 
\ee
Then, the ADM mass of the system is defined as the following
surface integral at radial infinity:
\be440
M_{ADM} = -{1 \over 8\pi G} \int_{\partial M} N(K - K_0)\ , 
\ee
where $N= e^{-V}$ is the norm of the time-like Killing vector and $K$
is the extrinsic curvature.  In our case it is given by
$K = n^r \partial_r A \sim e^{-W} \rho^2$, where $A$ is the asymptotic
area and $n^r = e^{-W}$ is the normal vector.  $K_0$ corresponds to $K$
defined in the same (reference) non-flat background but without any
matter fields.

Carrying out the procedure for our particular case 
we arrive at the following result: 
\be450
M_{ADM} = {{q_1+q_2+q_3}} + {3\over 2}{\mu}\ , 
\ee
where we have  taken the Newton's constant $G={\pi\over 4}$.

\subsection{Condition for the existence of horizons}

We now turn to the discussion of the global space-time structure of
the solution. 
In particular horizons appear at zeros of the function
$f$ (or $e^{-2V}$ in (\ref{420})). Hence, we have to look for
solutions of the following, effectively cubic equation for $x\equiv
r^2$:
\be460
x^2 f =
g^2\left( x^3 + A x^2 - Bx + q_1 q_2 q_3 \right) 
= 0\ ,
\ee
with
\be482
A\equiv \sum_{i=1}^3 q_i +{1\over g^2} \qquad , \qquad \  B\equiv {\mu\over
g^2}-\sum_{i>j=1}^3q_iq_j>0 \ .
\ee
Note, a necessary condition for having at least one zero of 
(\ref{460}) for $x>0$ is $B > 0$.
The extrema of (\ref{460}) ($ (x^2 f(x))' = 0$)
are at
\be490 
x_{\pm} =
\textstyle{1\over 3}A(-1\pm y)\ , \ \ y=1+z\equiv \sqrt{1+ {{3B}\over
A^2}}>1 \ .
\ee

Thus discarding the  extremum $x_-<0$,
a sufficient constraint to have at least one horizon is that:
\be500
x_{+}^2 f(x_{+}) \leq 0\ , 
\ee
or equivalently (employing (\ref{482}), (\ref{490})) and (\ref{500})):
\be502
 -2z^3-3z^2+C\leq 0 \qquad , \qquad
 C\equiv \Big({3\over A}\Big)^3 \prod_{i=1}^3 q_i \le 1 \ ,
 \ee
with equality sign corresponding to the case of coinciding inner and outer
horizons. Introducing $\varphi= {\rm arccos}(z-{1\over 2})$,
%
%
the inequality (\ref{502}) becomes:
\be504
\cos 3 \varphi = \cos\left[3 \arccos\left(1+ {{3B}/A^2})^{1/2} - {1 \over
2}\right)\right]\ge 2C -1  \  ,
\ee
%
where $A$, $B$ and $C$ are given in (\ref{482}) and (\ref{502}).  It
is straightforward to transcribe this inequality as  a lower bound on the
 value of 
$B$ or equivalently of $\mu$. For the well-defined classical solution
the charges $q_i$ and the gauge coupling $g$  are assumed to be in the range
$q_i>1> g^2$. 
The bound on $z$
becomes especially 
explicit in the two
limiting cases
(i) $g^2q_i \ll 1$
and thus  $C\ll 1$ and  (ii) $g^2q_i \gg 1$ and $q_i\sim q_2\sim q_3$, and
thus $C -1\ll
1$~\footnote{ 
The case (i) can be approximated by $\cos 3\varphi = - \cos(
3\varphi + \pi) = -1 + {1 \over 2}(3\varphi + \pi)^2 \pm ... $ and
therefore $\varphi = - {\pi\over 3} + {2 \over 3} \sqrt C \pm ...$ , while
the (ii) can be approximated by $\cos3 \varphi = 1 + {9 \over 2}
\varphi^2 \pm ... = 2C -1$ and therefore $\varphi = {2 \over 3}
\sqrt{C-1} \pm ... $ .}: 
\be515
\ba{rcll} 
z 
\geq z_{crit}&=&\sqrt{C\over
3}(1+{\cal O}(C)) \ , & C \ll 1 \ , \\ z 
\geq z_{crit}&=&
{1 \over 2} + {2 \over 9} (C-1) +{\cal O}\left((C-1)^2\right)
\ , \quad & C-1 \ll 1\ .  
\ea 
\ee
In the case of $C\ll 1$, i.e.  $g^2q_i\ll 1$,
 one then obtains the following explicit
bound:
\be517
\mu\ge\mu_{crit}=
2\sqrt{g^2\prod_{i=1}^3q_i}+ g^2\sum_{i>j=1}^3q_iq_j+{\cal
O}\left((g^2q_i)^{3/2}q_i\right)
 \ \ \ g^2q_i\ll 1 \ ,
\ee
while the  second limit $C\sim 1$, i.e. $q_i\sim q$, $g^2q\gg 1$, 
corresponds to
the following bound:
\be519
\mu\ge\mu_{crit}={27\over 4} g^2q^2 + {5\over 2} q + 
{\cal O}(g^{-2})
\ , \  \ \ q_i\sim q, \ \ g^2q\gg 1 \ \ .
\ee
Choosing $\mu$ large enough in order to comply with the inequality
(\ref{504}) (and more explicitly, with (\ref{517}) and (\ref{519}) in
the case of special limits) ensures that the $f$-function has two
positive and one negative zero and can be written in the form:
\be510
f = {g^2 (r^2 + r_0^2)(r^2 - r_-^2)(r^2 - r_+^2) \over r^4} \ ,
\ee
where $r_{\pm}$ denote respectively the outer and inner horizons. In the
extreme limit $r_+^2 \rightarrow r_-^2 \to x_+$ the two horizons
coincide and $\mu$ saturates the lower bound,
i.e. $\mu=\mu_{crit}$, as discussed above.  (For the discussion of an
equivalent bound for four-dimensional charged black holes with constant negative
cosmological constant we refer to
\cite{240}.)

\subsection{Bekenstein-Hawking entropy and Hawking temperature}

The Bekenstein-Hawking entropy is specified by the area of the outer
horizon, and thus it is a valid concept for the black holes with
regular horizons.  In particular we are interested in the entropy of
the solutions that saturate the bound on $\mu$, i.e. those with the
inner and outer horizon coinciding.

In  case (i) with $g^2q_i\ll 1$ the expression can be cast, by using  
(\ref{515}), (\ref{517})  and (\ref{100}), into the following explicit
form:
\be516
S_{crit}={A_{r_+}\over {4 G}}= 2\pi\sqrt{\prod_{i=1}^3q_i}\left(1+{\cal
O}((g^2q_i)^{1/2})\right)\ , \ \ \ g^2q_i\ll 1 \ .
\ee
The expression for the entropy resembles very closely that for the
BPS-saturated black holes in the ungauged supergravity
case~\cite{280}, except that the parameters $q_i$ are related to the
physical charges $\tilde q_i$ through equations (\ref{392}).
Notice, that the radius of the $AdS$ space scales with
the inverse gauge coupling $g$ and therefore this limit ($g^2 q_i \ll 1$)
corresponds to the leading order term in the large $N$ expansion ($N
\sim 1/g^2$) which, interestingly, is independent of $g$.

In  case (ii) the entropy assumes, using (\ref{515}), (\ref{519}) and
(\ref{100}), the following form:
\be526
S_{crit}=2\pi\sqrt{{27\over 8}q^3}\left(1+{\cal O}((g^2q)^{-1})\right) \ , \\ \
q_i\sim q\ , \ \ g^2q\gg 1 \ .
\ee
Note the new numerical factors in this entropy.
(It would  of course be interesting to obtain an explicit form for the  entropy 
$S_{crit}$  for the 
whole range of $g^2q_i$ values.)

As usual, the Hawking temperature is determined by the periodicity of
the Euclidean time. The $(r,t)$-part of the metric in the Euclidean
time $\tau$ is conformally equivalent to
\be520
dr^2 + e^{-6U} f^2  d\tau^2 \ .
\ee
For $r^2 \simeq r_+^2$  we have
\be530
e^{-3U} f = {2 g^2 (r_+^2 + r_0^2)(r_+^2 -r_-^2) \over 
\sqrt{ (r_+^2 +q_1)(r_+^2 +q_2)(r_+^2 +q_3)}}
\; (r - r_+) \ .
\ee
Therefore in order to cancel the conical singularity the
periodicity
\be540
\tau \sim \tau + {1 \over T_H} 
\ee
implies that the Hawking temperature is  given by
\be550
T_H = {g^2 (r_+^2 + r_0^2)(r_+^2 -r_-^2) \over \pi
\sqrt{(r_+^2 +q_1)(r_+^2 +q_2)(r_+^2 +q_3)}}
\ee
Thus, if both horizons coincide $r_- = r_+$ the Hawking temperature
vanishes, but the solution does not coincide with the BPS-saturated solution
\cite{130}. This result is analogous to the ``cold'' AdS black
holes discussed in \cite{220} in the context of $D=4$ gauged
supergravity.

Notice, for an uncharged black hole the situation is completely
different. In this case $H_1 = H_2 = H_3 =1$ and
one finds for the $f$-function
\be560
f = {g^2 \over r^2}\Big(r^2 - r_-^2 \Big)\Big(r^2 - r_+^2\Big)
\ee
with $r_{\pm}^2 = {1\over2 g^2} \big( -1 \pm \sqrt{1 + 4 g^2 \mu} \big)$.
In this case the Hawking temperature becomes $T_H = {g^2 \over \pi r_+}
(r_+^2 - r_-^2)$ and since $r_-^2$ is negative it can never vanish.
Instead, it diverges for $\mu \rightarrow 0, \infty$ and has
a minimum at $\mu = {3 \over 4 g^2}$: $T_H^{min} = {\sqrt 2 g \over \pi}$.
Therefore, at this temperature the black hole is in thermal
equilibrium with the thermal radiation (see discussion in \cite{250})
and it gives a lower bound for the black hole
size $r_+^{min} = 1/\sqrt{2 g^2}$.


\section{Conclusion}


In this paper we investigated charged black holes of $D=5$ gauged
$N$=2 supergravity, by deriving the complete set of equations of
motion (with a specific Ansatz for the fields in the theory).  In
order to keep these solutions as general as possible we considered
static solution with the spatial geometry not only of a three-sphere
$S_3$ ($k=1$), as it would be natural for a static black hole, but we
also included examples of Einstein spaces with constant spatial
curvatures $k=-1$ and $k=0$.

The Ans\"atze for the metric and the scalar fields are a natural
generalization of the solutions for the ungauged supergravity. The
main difference appears in an additional term in the non-extremality
function, which is due to the gauging of the theory, and it is a
proportional to the gauge coupling $g^2$ (see (\ref{170})). We showed,
that in the Einstein and scalar field equations this additional term
is precisely compensated by the contribution from the (gauged)
potential, thus rendering the form of these equations
to be the same as in the ungauged case. Therefore the static
spherically symmetric solutions of the ungauged supergravity can be
promoted to solutions of the gauged supergravity by adding to the
non-extremality function the specific term proportional to $g^2$ (see
(\ref{170})).  However, this additional term
has important consequences for the global space-time structure of
these solutions; in the BPS-saturated (supersymmetric) limit the
solutions have naked singularities!  As a consequence, there is a
lower bound on the non-extremality parameter $\mu$ (or equivalently
the ADM mass), determined by the condition that the Hawking
temperature vanishes, i.e.\ the outer and inner horizon coincide.

We demonstrated these results on a representative example of the $STU$
model, i.e.\ a three-charge black hole solution with two scalar fields
(which incidentally is also a solution of $D=5$ gauged $N=4$ and $N=8$
supergravity theory). In this case, the solution can be expressed
completely in terms of harmonic functions.  (For a general
prepotential, as discussed at the end of Section 3 it is not possible
to solve the equations of motion in terms of harmonic functions only.)
In particular we found the explicit form of the solution, derived the
$ADM$ mass and found the explicit bounds on the non-extremality
parameter that ensures regular horizons.  When the bound is saturated
the solution is not supersymmetric, i.e.  $\mu=\mu_{crit}\ne 0$,
however the Hawking temperature vanishes.  For this limit we
calculated also the entropy for the solution with $k=1$, which in the case of
small $g^2$ (i.e. large $N$ limit on the CFT side)  assumes a form 
that resembles that of the corresponding BPS-saturated solution in the ungauged
supergravity.  This example
may serve as an interesting gravity background for the study of the
AdS/CFT correspondence.

The analysis presented here for the case of $D=5$ gauged $N=2$
supergravity has a natural generalization to the case of non-extreme
black hole solutions of $D=4$ gauged supergravity. The same
qualitative changes between the solutions in the gauged and ungauged
cases~\cite{140} are expected to take place.  Namely, only the
non-extremality function is expected to be modified by a specific term
proportional to $g^2$, while other fields would satisfy the same
equations of motion as in the ungauged case. The global space-time is
expected to changed accordingly; the regular solutions again have to
satisfy a lower bound on the non-extremality parameter $\mu\ge
\mu_{crit}$.  A representative solution to demonstrate these phenomena
explicitly would again be within the $STU$ model, corresponding to the
four-charge static black hole solutions.

\vskip 0.5cm
\noindent{\bf Acknowledgments}
\vskip 0.3cm

We would like to thank A. Chamseddine and T. Mohaupt for useful
discussions and correspondence. M.C. would like to thank the
organizers of the 32nd International Symposium Ahrenshoop on the
Theory of Elementary Particles, where part of the work was completed,
for hospitality. K.B. would like thank the Dept.\ of Physics of the
Univ.\ of Pennsylvania for hospitality.  The work is supported by the
Department of energy grant no. DOE-FG02-95ER40893, by the
University of Pennsylvania Research Foundation grant (M.C.)
and by the DFG (K.B.).




%
%


\providecommand{\href}[2]{#2}\begingroup\raggedright\endgroup


\end{document}